\documentclass[%
 aip,
 jmp,%
 amsmath,amssymb,
 reprint,%
]{revtex4-1}

\usepackage{subfigure}
\usepackage{graphicx}
\graphicspath{{./}{./images/}{./images_all/}}
\DeclareGraphicsExtensions{.eps}

\usepackage{dcolumn}
\usepackage{bm}

\begin{document}

\preprint{AIP/123-QED}

\title[APPLIED PHYSICS LETTERS xx (2011)]{Hybrid spintronics and straintronics: 
A magnetic technology for ultra low energy computing and signal processing }

\author{Kuntal Roy}
\email{royk@vcu.edu.}
\author{Supriyo Bandyopadhyay}
\affiliation{Department of Electrical and Computer Engineering, Virginia Commonwealth University, Richmond, VA 23284, USA}
\author{Jayasimha Atulasimha}
\affiliation{Department of Mechanical and Nuclear Engineering, Virginia Commonwealth University, Richmond, VA 23284, USA}

\date{\today}

\begin{abstract}

The authors show that the magnetization of a magnetostrictive/piezoelectric multiferroic single-domain 
shape-anisotropic nanomagnet 
can be switched with very small voltages that generate strain in the magnetostrictive layer. 
This can be the basis of  ultralow
power computing and signal processing. 
With appropriate material choice, 
the energy dissipated per switching event can be reduced to $\sim$45 $kT$ at room temperature for a
switching delay of $\sim$100 ns and $\sim$70 $kT$ for a switching delay of $\sim$10 ns, if the 
energy barrier separating the
 two stable magnetization directions  is $\sim$32 $kT$. 
Such devices can be powered by harvesting energy exclusively from the environment without the need for a battery.   
\end{abstract}

\maketitle

The primary obstacle to continued downscaling of digital electronic devices in accordance with Moore's
 law is the excessive energy dissipation that takes place in the device during switching of bits. 
 Every charge-based device (e.g. MOSFET) has a 
fundamental shortcoming in this regard. They are switched by injecting or extracting an amount of charge 
$\Delta Q$ from the 
device's active region  
with a potential gradient $\Delta V$, leading to an inevitable energy dissipation of $\Delta Q \times \Delta V$. 
Spin based devices, on the other hand, are switched by flipping spins without moving any charge in space ($\Delta Q$ = 0)
and
causing a current flow. 
Although some energy is still dissipated in flipping spins, it can be considerably less than the energy $\Delta Q \times \Delta V$ associated 
with current flow. This gives ``spin'' an advantage over ``charge'' as a state variable to encode digital bits.

Recently, it has been shown that the {\it minimum} energy dissipated to switch a charge-based device like a transistor 
at a temperature $T$ is $\sim$$NkTln(1/p)$, 
where $N$ is the number of information carriers (electrons or holes) in the device and $p$ is the bit error probability 
\cite{salahuddin}. 
On the other hand, the minimum energy dissipated  to switch a single domain nanomagnet (which is a collection of $M$ spins) 
can be only $\sim$$kTln(1/p)$  since the exchange interaction between the spins makes all of them rotate together in unison 
like a giant classical spin \cite{salahuddin,cowburn}. This gives the magnet an advantage over the transistor.

Unfortunately, the magnet's advantage is lost  if the 
method adopted to switch it is so inefficient that the  energy dissipated in the switching circuit
far exceeds the energy dissipated in the magnet. Regrettably, this is often the case. 
A magnet is usually flipped with either a magnetic field generated by a current \cite{alam}, or
a spin polarized current exerting either a spin transfer torque \cite{ralph} or causing domain wall motion 
\cite{ohno}.  The 
energy dissipated  to switch a magnet with current-generated magnetic field was reported in ref. [3]  
as 10$^{11}$ - 10$^{12}$ $kT$ for a 
switching delay of $\sim$1 $\mu$s, which clearly makes it 
impractical. On the other hand, the energy
dissipated to switch a two-dimensional nanomagnet in  $\sim$1 ns with spin transfer torque is estimated to be
$\sim$2$\times$10$^8$ $kT$  if the energy barrier separating the two stable magnetization orientations is 32 $kT$
(so that the equilibrium bit error probability $p$ = $e^{-32}$) \cite{kuntal}.
 Therefore, both of these switching methods 
are very energy-inefficient since the switching circuit dissipates far more energy than the minimum $\sim$$kTln(1/p)$ needed. 
In fact, they are so inefficient that they might not even make the magnet superior
to the transistor which can be switched in sub-ns while dissipating 10$^7$ - 10$^8$ $kT$ of
energy in a circuit \cite{comment}.  
Only domain wall motion has been shown to  be a relatively energy-efficient  
switching mechanism since there is at least one report of switching a 
nanomagnet in 2 ns while dissipating
10$^4$ - 10$^5$ $kT$ of energy \cite{ishawata}. Thus, there is a need to identify energy-efficient mechanisms 
for switching a magnet. This is the motivation for 
this work.

Recently, we have shown that the magnetization of a strain-coupled 
piezoelectric/magnetostrictive {\it multiferroic} nanomagnet can be switched by
stressing the magnetostrictive layer with a small voltage applied to the piezoelectric layer \cite{our_apl}. 
Such multiferroic systems have now become commonplace \cite{ramesh,brittlinger,nature} and there are
proposals for using them in magnetic logic and memory \cite{our_apl,fashami,kuntal,wolf}. In this method, 
an electrostatic potential applied to the piezoelectric layer of a multiferroic magnet generates in it a strain that
is elastically transferred to the magnetostrictive layer if the latter layer is considerably thinner. This stresses the magnetostrictive 
layer and 
causes its magnetization to rotate. Such rotations have been demonstrated experimentally \cite{brittlinger}.
\begin{figure}
\includegraphics[width=2.5in]{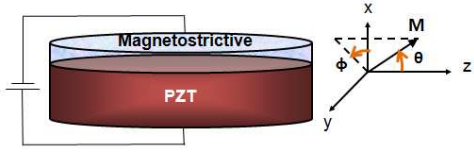}
\caption{\label{fig:multiferroic} An elliptical multiferroic nanomagnet stressed with an applied voltage.}
\end{figure}
Consider an ellipsoidal multiferroic magnet with uniaxial shape anisotropy as shown in Fig. 1. 
 The magnetostrictive
layer is assumed to be 10 nm thick and the piezoelectric layer is 40 nm thick, which ensures that most 
of the strain
generated in the piezoelectric layer by an applied voltage is transferred to the magnetostrictive layer.
We assume that the piezoelectric layer is lead-zirconate-titanate (PZT) and the magnetostrictive layer is 
polycrystalline nickel or cobalt or Terfenol-D. For Terfenol-D, the major axis is assumed to be $\sim$102 nm and the minor 
axis is $\sim$98 nm.
Because of shape anisotropy, the two magnetization orientations parallel
to the easy axis (major axis of the ellipse, or the z-axis) are stable and store the binary bits 0 and 1. 
The energy barrier between these two orientations (i.e. the shape anisotropy barrier) will be 0.8 eV or $\sim$32 $kT$
at room temperature for the Terfenol-D system, which makes the equilibrium bit error probability $e^{-32}$.
Let us assume that the magnetization is initially oriented along the -z-axis. Our task is to switch the nanomagnet 
 so that the final orientation is along the +z-axis. 
We do this by applying a stress along the easy axis (z-axis). Since the stress is generated with a voltage $V$ applied to the piezoelectric layer, the energy dissipated during 
turn-on is $(1/2)CV^2$ while that dissipated during turn-off is $(1/2)CV^2$, where $C$ is the capacitance
of the piezoelectric layer, and we have assumed that the voltage is applied abruptly or non-adiabatically.

There is an additional dissipation $E_d$ in the nanomagnet due to Gilbert damping \cite{datta},
which is of the order of the energy barrier modulation in the nanomagnet, i.e. by how much the energy barrier between the two stable
magnetizations is lowered to switch between the states.
The total energy dissipated in the 
switching process is therefore $E_{total} = CV^2 + E_d$. Thus, in order to calculate 
$E_{total}$ as a function of switching delay, we have to calculate four quantities:
 (1) the stress needed to switch the magnetization within the given
delay, (2) the voltage $V$ needed to generate this stress, (3) the capacitance $C$ of the multiferroic, and (4) $E_d$
which is calculated by following the prescription of ref. [12] (see the supplementary material).

In order to find the stress $\sigma$ required to switch a magnetostrictive  nanomagnet in a given time delay $\tau$, we solve the Landau-Lifshitz-Gilbert (LLG)
equation for a single domain magnetostrictive nanomagnet subjected to stress $\sigma$. We then relate $\sigma$ to the strain $\epsilon$ 
in 
the nanomagnet from Hooke's law ($\epsilon = \sigma/Y$, where $Y$ is the Young's modulus of the nanomagnet) and 
find the voltage $V$ that generates that strain in the PZT layer based on the $d_{31}$ coefficient of PZT and its thickness. 
Finally, we calculate the capacitance of the multiferroic system by treating it as a parallel-plate capacitor while 
taking the relative dielectric constant of PZT to be 1000. This allows us to find the energy dissipated in the switching 
circuit ($CV^2$) 
as a function of the switching delay $\tau$.

In the supplementary material accompanying this letter, we show that both
stress and shape anisotropy act like a torque on the magnetization of the nanomagnet.
This torque per unit volume of the nanomagnet is given by 
\begin{equation}
\mathbf{T_E} (t) = - \mathbf{n_m}(t) \times \nabla E[\theta(t),\phi(t)],
\end{equation}
where $E[\theta(t),\phi(t)]$ is the total energy of the nanomagnet at an instant of time $t$. It is the
sum of shape anisotropy energy and stress anisotropy energy, both of which depend on the magnetization
orientation at the given instant. We adopt the spherical coordinate system whereby the magnetization
is along the radial direction so that its orientation at any instant is specified by the instantaneous 
polar angle $\theta(t)$ and the azimuthal angle $\phi(t)$. 

In the supplementary material, we show that we can write the torque as 
\begin{eqnarray}
\mathbf{T_E} (t) &=&  - \{2 B (\phi(t)) sin\theta(t) cos\theta(t)\} \mathbf{\hat{e}_\phi} \nonumber \\
&&  - \{B_{0e}(\phi(t)) \, sin\theta (t)\} \mathbf{\hat{e}_\theta},
\label{stress-torque}
\end{eqnarray}
where $\bf{\hat{e}_\theta}$ and $\bf{\hat{e}_\phi}$ are unit vectors in the $\theta$- and $\phi$-directions, and
\begin{subequations}
\begin{align}
B_0(\phi(t)) &= \cfrac{\mu_0}{2} \, M_s^2 \Omega \left\lbrack N_{xx} cos^2\phi(t) + 
N_{yy} sin^2\phi(t) - N_{zz}\right\rbrack \\
B_{stress} &= (3/2) \lambda_s \sigma \Omega \\
B(\phi(t)) &= B_0(\phi(t)) + B_{stress} \\
B_{0e}(\phi(t)) &= \cfrac{\mu_0}{2} \, M_s^2 \Omega (N_{xx}-N_{yy}) sin(2\phi(t)).
\end{align}
\end{subequations} 
Here $M_s$ is the saturation magnetization of the nanomagnet, $\Omega$ is its volume, $\mu_0$
is the permeability of free space, $\lambda_s$ is the 
magnetostrictive coefficient of the magnetostrictive layer,  and $N_{\beta \beta}$ is the demagnetization 
factor in the $\beta$ direction, which can be calculated from the shape and size of the nanomagnet 
[see the supplementary material].

The magnetization dynamics of the single-domain nanomagnet  is described by the
 Landau-Lifshitz-Gilbert (LLG) equation as follows.
\begin{equation}
\cfrac{d\mathbf{n_m}(t)}{dt} + \alpha \left(\mathbf{n_m}(t) \times \cfrac{d\mathbf{n_m}(t)}
{dt} \right) = \cfrac{\gamma}{M_V} \mathbf{T_E}(t) 
\label{LLG}
\end{equation}
where $\mathbf{n_m}(t)$
is the nomalized magnetization, $\alpha$ is the dimensionless phenomenological Gilbert damping constant, 
$\gamma = 2\mu_B \mu_0 /\hbar$ is the gyromagnetic ratio for electrons, and $M_V=\mu_0 M_s \Omega$. 

From this equation, we can derive two coupled equations that describe the $\theta$- and $\phi$-dynamics.
The derivation can be found in the supplementary material. The final result is:
\begin{eqnarray}
\left(1+\alpha^2 \right) \theta'(t) &=& -\frac{\gamma}{M_V} \lbrack B_{0e}(\phi(t)) sin\theta(t) \nonumber \\
&& + 2\alpha B (\phi(t)) sin\theta (t)cos\theta (t) \rbrack, \label{eq:theta_dynamics} \\
\left(1+\alpha^2 \right) \phi'(t) &=& \frac{\gamma}{M_V} \left\lbrack \alpha B_{0e}(\phi(t)) - 2 B(\phi(t)) cos\theta(t) \right\rbrack  \nonumber\\
&& \qquad\qquad\qquad (sin\theta(t) \neq 0).
\label{eq:phi_dynamics}
\end{eqnarray}

Clearly, the $\theta$- and $\phi$-motions are coupled and hence these equations have to be solved numerically.
We assume that the initial orientation of the nanomagnet is close to the --z-axis ($\theta$ = 179$^{\circ}$). 
It cannot be {\it exactly} along the --z-axis ($\theta$ = 180$^{\circ}$) since then the torque acting on it will be 
zero (see Equation \eqref{stress-torque}) and the magnetization will never rotate under any stress. Similarly, we cannot
make the final state align exactly along the +z-axis ($\theta$ = 0$^{\circ}$) since there too the torque vanishes.
 Hence we assume that the final state is 
$\theta$ = 1$^{\circ}$. Thus, both initial and final states are 1$^{\circ}$ off from the easy axis.Thermal
fluctuations can easily deflect the magnetization by 1$^{\circ}$.

\begin{figure}
\includegraphics[width=2.4in]{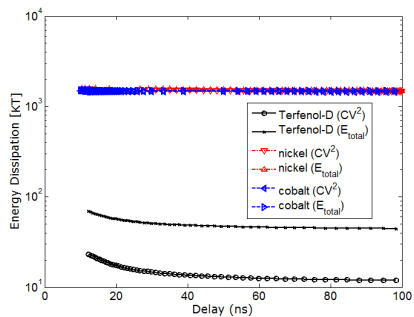}
\caption{\label{fig:energy_harvesting_delay_energy_bothCV2andEtotal} Energy dissipated in the switching circuit 
($CV^2$) and the total energy dissipated ($E_{total}$) as functions of delay for three
different materials used as the magnetostrictive layer in the multiferroic nanomagnet.}
\end{figure}
 
We apply the voltage generating stress abruptly at time $t$ = 0. This rotates the magnetization
away from near the easy axis ($\theta=179^{\circ}$) since the latter is no longer the minimum energy
state. The new energy minimum is at $\theta = 90^{\circ}$. We maintain the stress until $\theta$ reaches 90$^{\circ}$ which places
the magnetization approximately along the in-plane hard axis (y-axis). Then we reduce the voltage to zero abruptly. Subsequently,
shape anisotropy takes over and the magnetization vector will rotate towards the easy axis since that becomes the minimum 
energy state. The question is which direction along the easy axis will the magnetization vector relax to. Is it the --z-axis at $\theta = 179^{\circ}$ (wrong state)
or the +z-axis at $\theta = 1^{\circ}$ (correct state)? That is determined by the sign of  $B_{0e}(\phi(t))$ when $\theta$
reaches 90$^{\circ}$. If $\phi$ at that instant is less than 90$^{\circ}$, then $B_{0e}(\phi(t))$ is
positive which makes the time derivative of $\theta$ negative (see Equation~\eqref{eq:theta_dynamics}) so that $\theta$
continues to {\it decrease} and the magnetization reaches the correct state close to the +z-axis. The coupled 
$\theta$- and $\phi$-dynamics {\it ensures} that this is the case as long as the stress exceeds a minimum value. 
Thus, successful switching requires  a minimum
stress.

Once we have found the switching delay $\tau$ for a given stress $\sigma$ by solving Equations (\ref{eq:theta_dynamics}) and 
(\ref{eq:phi_dynamics}), we can invert the relationship to find $\sigma$ versus $\tau$ and hence the energy dissipated
versus $\tau$. This is shown in Fig. 2 where we plot the energy dissipated in the switching circuit ($CV^2$),
as well as the total energy dissipated  ($E_{total}$) versus delay for three materials 
chosen for the magnetostrictive layer. For Terfenol-D, the stress required to switch in 100 ns  
is 1.92 MPa and that required to switch in 10 ns is 2.7 MPa.

Note that for a stress of 1.92 MPa, the stress anisotropy energy $B_{stress}$
is 32.7 $kT$  while for 2.7 MPa, it is 46.2 $kT$. As expected, they are larger than the shape anisotropy barrier of $\sim$32 $kT$ which 
had to be overcome by stress to switch.  A larger excess energy is needed to switch faster. The energy dissipated and lost as 
heat in the switching circuit ($CV^2$)
is only 12 $kT$ for a delay of 100 ns and 23.7 $kT$ for a delay of 10 ns. The total energy dissipated is 45 $kT$ for a delay of
100 ns and 70 $kT$ for a delay of 10 ns. Note that in order to increase the switching speed by a factor of 10, the dissipation needs to
increase by  a factor of 1.6. Therefore, dissipation increases {\it sub-linearly} with speed, which bodes well for energy efficiency.

With a nanomagnet density of 
10$^{10}$ cm$^{-2}$ in a memory or logic chip, the dissipated power density would have been only 2 mW/cm$^2$ to switch in 100 ns
and 30 mW/cm$^2$ to switch in 10 ns, if 10\% of 
the magnets switch at any given time (10\% activity level). Note that unlike transistors, magnets 
have {\it no leakage} and no standby power dissipation, which is an important additional benefit. 

Such {\it extremely low power} and yet {\it high density} magnetic logic and 
memory systems, composed of multiferroic nanomagnets, can be powered by existing energy harvesting systems 
\cite{roundy, anton, lu, jeon} that harvest energy from the environment without the need for an external battery. 
These processors are uniquely suitable for implantable medical devices, e.g. those implanted in a patient's brain 
that monitor brain signals to warn of impending epileptic seizures. They can run on energy harvested from the 
patient's body motion. For such 
applications, 10-100 ns switching delay is adequate. Speed is not the primary concern, but energy
dissipation is. These hybrid spintronic/straintronic processors 
can be also incorporated in
``wrist-watch'' computers  powered by  arm movement, 
buoy-mounted computers for tsunami monitoring (or naval applications) that harvest energy from sea waves, 
or structural health monitoring systems for bridges and buildings that are powered solely by mechanical 
vibrations due to wind or passing traffic.


\end{document}


\maketitle

In this supplementary section, we first derive the equations describing the time evolution
of the polar angle $\theta(t)$ and the azimuthal angle $\phi(t)$ of the magnetization vector.
We do this starting from the Landau-Lifshitz-Gilbert (LLG) equation.

\section{Magnetization dynamics of a multiferroic nanomagnet: Solution of the Landau-Lifshitz-Gilbert equation}

Consider an isolated nanomagnet of ellipsoidal shape lying in the y-z plane with its major axis aligned along the z-direction and minor axis along the y-direction. The dimension of the major axis is $a$ and that of the minor axis is
$b$, while the thickness is $l$. The volume of the nanomagnet is $\Omega=(\pi/4)a b l$. 
 Let $\theta(t)$ be the angle subtended by the magnetization with the +z-axis at any instant of time $t$ and $\phi(t)$ be the angle 
subtended by the projection of the magnetization vector on the x-y plane with the +x
axis. We call $\theta(t)$ the polar angle and $\phi(t)$ the azimuthal angle. These are 
represented in Fig. 1 of the main paper.

The total energy of the single-domain nanomagnet is the sum of the uniaxial shape 
anisotropy energy $E_{SHA}$ and  the stress anisotropy energy $E_{STA}$:
\begin{equation}
E = E_{SHA} + E_{STA},
\end{equation}
where 
\begin{equation}
E_{SHA} = (\mu_0/2) M_s^2 \Omega N_d,
\end{equation}
with $M_s$ being the saturation magnetization and $N_d$  the demagnetization factor expressed as 
\begin{equation}
N_d = N_{zz} cos^2\theta(t) + N_{yy} sin^2\theta(t) \, sin^2\phi(t) + N_{xx} sin^2\theta(t) \, cos^2\phi(t)
\end{equation}
Here $N_{zz}$, $N_{yy}$, and $N_{xx}$ are the components of $N_d$ along the $z$-axis, 
$y$-axis, and $x$-axis, respectively. If $l \ll a,b$, then $N_{zz}$, $N_{yy}$, and $N_{xx}$ are given by~\cite{RefWorks:157}
\begin{subequations}
\begin{eqnarray}
		N_{zz} &=& \frac{\pi}{4} \left(\frac{l}{a} \right) 
\left\lbrack 1 - \frac{1}{4}\left(\frac{a-b}{a} \right) - \frac{3}{16}
\left(\frac{a-b}{a} \right)^2 \right\rbrack \\
		N_{yy} &=& \frac{\pi}{4} \left(\frac{l}{a} \right) 
\left\lbrack 1 + \frac{5}{4}\left(\frac{a-b}{a} \right) + \frac{21}{16}
\left(\frac{a-b}{a} \right)^2 \right\rbrack\\
	  N_{xx} &=& 1 - (N_{yy} + N_{zz}).
\end{eqnarray}
\end{subequations}
which shows that $N_{xx} \gg N_{yy}, N_{zz}$.

Note that in the absence of any stress, uniaxial shape anisotropy will favor 
lining up the magnetization along the major axis ($z$-axis) [$\theta$ = 0, $\phi$ = 90$^{\circ}$] by minimizing $E_{SHA}$, which is why we will call the major axis the ``easy axis'' and the minor axis (y-axis) the ``hard axis''. 
We will assume that a force along the $z$-axis (easy axis) generates stress in the magnet.
 In that case, 
the stress anisotropy energy is given by
\begin{equation}
E_{STA} = - (3/2) \lambda_s \sigma \Omega \, cos^2\theta(t),
\end{equation}
where $(3/2) \lambda_s$ is the magnetostriction coefficient of the nanomagnet and 
$\sigma$ is the stress. Note that a positive $\lambda_s \sigma$ product will favor alignment of the magnetization along the 
major axis (z-axis), while a negative $\lambda_s \sigma$ product will favor alignment along the minor axis ($y$-axis), because that will minimize $E_{STA}$. In our convention, a compressive stress is negative and tensile stress is positive. Therefore, in a material like Terfenol-D that has positive $\lambda_s$, a compressive stress will favor alignment along the minor axis, and tensile along the major axis. The situation will be opposite with nickel that has negative $\lambda_s$.
 
At any instant of time, the total energy of the nanomagnet can be expressed as 
\begin{equation}
E(t) = E[\theta(t), \phi(t)] = B(\phi(t)) sin^2\theta(t) + C
\end{equation}
where 
\begin{subequations}
\begin{align}
B_0(\phi(t)) &= \cfrac{\mu_0}{2} \, M_s^2 \Omega \left\lbrack N_{xx} cos^2\phi(t) + 
N_{yy} sin^2\phi(t) - N_{zz}\right\rbrack \\
B_{stress} &= (3/2) \lambda_s \sigma \Omega \displaybreak[3]\\
B(\phi(t)) &= B_0(\phi(t)) + B_{stress} \\
C &= \cfrac{\mu_0}{2} M_s^2 \Omega N_{zz} - (3/2) \lambda_s \sigma \Omega.
\end{align}
\end{subequations}
Note that $B_0(\phi(t))$ is always positive, but $B_{stress}$ can be negative or positive according to the sign of the $\lambda _s \sigma$ product.

The magnetization \textbf{M}(t) of the magnet has a constant magnitude at any given temperature but a variable direction,
so that we can represent it by the vector of unit norm 
$\mathbf{n_m}(t) =\mathbf{M}(t)/|\mathbf{M}| = \mathbf{\hat{e}_r}$ where 
$\mathbf{\hat{e}_r}$ is the unit vector in the radial direction in spherical coordinate 
system represented by ($r$,$\theta$,$\phi$). The other two unit vectors in the spherical 
coordinate system are denoted by $\mathbf{\hat{e}_\theta}$ and $\mathbf{\hat{e}_\phi}$ 
for $\theta$ and $\phi$ rotations, respectively. Note that
\begin{eqnarray}
{\bf \nabla} E(t) = {\bf \nabla} E[\theta(t),\phi(t)] &=& \cfrac{\partial E(t)}{\partial \theta(t)} \, \mathbf{\hat{e}_\theta} + \cfrac{1}{sin\theta(t)} \,\cfrac{\partial E(t)}{\partial \phi(t)} \, \mathbf{\hat{e}_\phi} \\
\cfrac{\partial E(t)}{\partial \theta(t)} &=& 2 B sin\theta (t) cos\theta(t) \\
\cfrac{\partial E(t)}{\partial \phi(t)} &=& -\frac{\mu_0}{2} \, M_s^2 \Omega (N_{xx}-N_{yy}) sin(2\phi(t))
 sin^2\theta (t) = - B_{0e}(\phi(t)) \, sin^2\theta (t) \nonumber \\
\label{eq:stress-torque2}
\end{eqnarray}
where $B_{0e}(\phi(t))=\cfrac{\mu_0}{2} \, M_s^2 \Omega (N_{xx}-N_{yy}) sin(2\phi(t))$. The torque acting on the magnetization within unit volume due to shape and stress anisotropy is
\begin{eqnarray}
\mathbf{T_E} (t) &=& - \mathbf{n_m}(t) \times \nabla E[\theta(t),\phi(t)] \nonumber\\
								 &=& - \mathbf{\hat{e}_r} \times \lbrack \{2 B(\phi(t)) sin\theta(t) cos\theta(t)\} \mathbf{\hat{e}_\theta} - \{B_{0e}(\phi(t))\,  sin\theta (t)\} \mathbf{\hat{e}_\phi} \rbrack   \nonumber\\
								 &=& - \{2 B (\phi(t)) sin\theta(t) cos\theta(t)\} \mathbf{\hat{e}_\phi} - \{B_{0e} (\phi(t)) \, sin\theta (t)\} \mathbf{\hat{e}_\theta}
\label{stress-torque}
\end{eqnarray}
\noindent
 
The magnetization dynamics of the single-domain magnet under the action of various torques is described by the
 Landau-Lifshitz-Gilbert (LLG) equation as follows.
\begin{equation}
\cfrac{d\mathbf{n_m}(t)}{dt} + \alpha \left(\mathbf{n_m}(t) \times \cfrac{d\mathbf{n_m}(t)}
{dt} \right) = \cfrac{\gamma}{M_V} \mathbf{T_E}(t) 
\label{LLG}
\end{equation}
where $\alpha$ is the dimensionless phenomenological Gilbert damping constant, $\gamma = 2\mu_B/\hbar$ is the gyromagnetic ratio for electrons and is given by $2.21\times 10^5$ (rad.m).(A.s)$^{-1}$, and $M_V=\mu_0 M_s \Omega$. In the spherical coordinate system, 
\begin{equation}
\cfrac{d\mathbf{n_m}(t)}{dt} = \theta '(t) \, \mathbf{\hat{e}_\theta} + sin \theta (t)\, \phi '(t) 
\,\mathbf{\hat{e}_\phi}.
\end{equation}
where the prime denotes first derivative with respect to time. Accordingly,
\begin{equation}
\alpha \left(\mathbf{n_m}(t) \times \cfrac{d\mathbf{n_m}(t)}{dt} \right) = - \alpha sin 
\theta(t) \, \phi'(t) \,\mathbf{\hat{e}_\theta} +  \alpha \theta '(t) \, \mathbf{\hat{e}_\phi}
\end{equation}
and 
\begin{equation}
\cfrac{d\mathbf{n_m}(t)}{dt} + \alpha \left(\mathbf{n_m}(t) \times \cfrac{d\mathbf{n_m}(t)}
{dt} \right) = (\theta '(t) - \alpha sin \theta (t)\, \phi'(t)) \, \mathbf{\hat{e}_\theta} 
+ 
(sin \theta(t) \, \phi ' (t) + \alpha \theta '(t)) \,\mathbf{\hat{e}_\phi}.
\end{equation}
Equating the $\hat{e}_\theta$ and $\hat{e}_\phi$ components in both sides of Equation
(\ref{LLG}), we get
\begin{subequations}
\begin{eqnarray}
\theta ' (t) - \alpha sin \theta(t) \, \phi'(t)  &=&  -\frac{\gamma}{M_V} \, B_{0e}(\phi(t)) \, sin\theta(t) \\
sin \theta (t) \, \phi '(t) + \alpha \theta '(t) &=& - \frac{\gamma}{M_V} 2 B(\phi(t)) sin\theta(t) cos\theta(t). 
\end{eqnarray}
\end{subequations}

Simplifying the above, we get
\begin{eqnarray}
\left(1+\alpha^2 \right) \theta'(t) &=& -\frac{\gamma}{M_V} \left\lbrack B_{0e}(\phi(t)) sin\theta(t) + 2\alpha B(\phi(t)) sin\theta (t)cos\theta (t) \right\rbrack \label{eq:theta_dynamics} \\
\left(1+\alpha^2 \right) \phi'(t) &=& \frac{\gamma}{M_V} \left\lbrack \alpha B_{0e}(\phi(t)) - 2 B(\phi(t)) cos\theta(t) \right\rbrack \quad (sin\theta(t) \neq 0).
\label{eq:phi_dynamics}
\end{eqnarray}

We will assume that the initial orientation of the magnetization is aligned close to the --$z$-axis 
so that $\theta_{initial} = 180^{\circ} - \epsilon$. If $\epsilon = 0$ and the magnetization is 
exactly along the easy axis [$\theta$ = 0$^{\circ}$, 180$^{\circ}$], then no amount of stress can budge 
it since the effective torque exerted on the magnetization by stress will be exactly zero
(see \eqref{stress-torque}). Such locations are called ``stagnation points''. 
Therefore, we will assume that $\epsilon = 1^{\circ}$. 
This is not an unreasonable assumption since thermal 
fluctuations can dislodge the magnetization from the easy axis and make $\epsilon \rightarrow 1^{\circ}$.

We should notice from Equation~\eqref{eq:theta_dynamics} that there is the possibility of one more stagnation point at 
$\theta(t)=\phi(t)=90^\circ$ 
[in-plane hard axis] since there $\theta'(t)=0$. At $\theta=90^\circ$,  Equation~\eqref{eq:theta_dynamics} becomes
\begin{equation}
\left(1+\alpha^2 \right) \theta'(t) = -\frac{\gamma}{M_V} \, B_{0e}(\phi(t)) = -\frac{\gamma}{M_V} \, \cfrac{\mu_0}{2} \, M_s^2 \Omega (N_{xx}-N_{yy}) sin(2\phi(t))
\end{equation}

\noindent
which indicates that as long as $\phi(t) < 90^\circ$, the magnetization vector  will continue to rotate 
towards the correct final state  without being stuck at $\theta$ = 90$^{\circ}$. 
This will avoid stagnation. Note that when $\theta$ = 90$^{\circ}$, we will stagnate if $\phi(t) = 90^\circ$,
rotate back towards the intial state along the --z-axis (wrong state) if $\phi(t) > 90^\circ$, and 
rotate towards the correct state along the +z-axis if $\phi(t) < 90^\circ$.

At high enough stress, 
the out-of-plane excursion of the magnetization vector is significant and $\phi(t) < 90^\circ$ so that
stagnation is indeed avoided and the correct state is invariably reached. However, at low stress,  the first term in 
Equation~\eqref{eq:phi_dynamics} will suppress out-of-plane excursion of the 
magnetization vector and try to constrain it to the nanomagnet's plane, thereby making $\phi(t)=90^\circ$.
This will result in stagnation when $\theta$ reaches 90$^{\circ}$ and switching will fail. 
Whether this happens or not depends on the relative strengths of the two terms in Equation~\eqref{eq:phi_dynamics} 
that counter each other. We need to avoid such low stresses to ensure successful switching. Thus, there is a 
minimum value of stress for which switching takes place. This minimum value is determined by material parameters.

One other issue deserves mention. We have shown explicitly that we can switch from an initial state close to the 
--z-axis to a final state close to the +z-axis. Can we do the opposite and switch from +z-axis to --z-axis? For a single
isolated magnet, this is always possible and the dynamics is identical. 
In magnetic random access memory (MRAM) systems, there are {\it two} strongly
dipole coupled magnet in close proximity, where one is the soft magnetic layer and the other is the hard magnetic layer.
In that scenario, there is a difference between switching from the anti-parallel to the parallel and from the parallel to 
the anti-parallel
arrangement of the two magnets owing to dipole coupling which is different in the two cases. 
This is not an issue for the single isolated magnet considered 
here. 

\section{Material parameters}

The material parameters that are used in the simulation are given in the 
Table~\ref{tab:material_parameters}~\cite{RefWorks:179,RefWorks:176,RefWorks:178,RefWorks:213, RefWorks:172, materials}.
They ensure that the shape anisotropy energy barrier is $\sim$32 $kT$.

\begin{table}[htbp]
\centering
\begin{tabular}{c||c|c|c}
& Terfenol-D & Nickel & Cobalt\\
\hline \hline
Major axis (a) & 101.75 nm & 105 nm & 101.75 nm\\
\hline
Minor axis (b) & 98.25 nm & 95 nm & 98.25 nm\\
\hline
Thickness (t) & 10 nm & 10 nm & 10 nm\\
\hline
Young's modulus (Y) & 8$\times$10$^{10}$ Pa & 2.14$\times$10$^{11}$ Pa & 2.09$\times$10$^{11}$ Pa \\
\hline
Magnetostrictive coefficient ($(3/2)\lambda_s$) & +90$\times$10$^{-5}$ & -3$\times$10$^{-5}$  & -3$\times$10$^{-5}$ \\
\hline
Saturation magnetization ($M_s$) & 8$\times$10$^5$ A/m & 4.84$\times$10$^5$ A/m & 8$\times$10$^5$ A/m \\
\hline
Gilbert's damping constant ($\alpha$) & 0.1 & 0.045 & 0.01 \\
\hline\hline
\end{tabular}
\caption{\label{tab:nickel}Material parameters for different materials.}
\label{tab:material_parameters}
\end{table}

\section{Procedure for determining the voltage required to generate a given stress in a magnetostrictive 
material}

In order to generate a stress $\sigma$  in a magnetostrictive layer, the strain in that material must 
be $\varepsilon = \sigma/Y$, where $Y$ is the Young's modulus of the material. We will assume that a voltage
applied to the PZT layer strains it and since the PZT layer is much thicker than the magnetostrictive layer,
all the strain generated in the PZT layer is transferred completely to the magnetostrictive layer. Therefore,
the strain in the PZT layer must also be $\varepsilon$. The electric field needed to generate this strain is calculated from
the piezoelectric coefficient $d_{31}$ of PZT  ($d_{31}=1.8\times10^{-10}$ m/V~\cite{pzt}) and the corresponding 
voltage is found by multiplying this field with the thickness of the PZt layer.

\section{Calculation of the energy $E_d$ dissipated internally within the magnet due to Gilbert damping}

Because of Gilbert damping in the magnet, an additional energy $E_d$ is dissipated when the magnet switches.
This energy is given by the expression 
\begin{equation}
\int_0^{\tau}P_d(t) dt ,
\end{equation}
where $\tau$ is the switching delay and $P_d(t)$, the dissipated power is given by \cite{RefWorks:319,RefWorks:124}
\begin{equation}
P_d(t) = \cfrac{\alpha \, \gamma}{(1+\alpha^2) \mu_0 M_s \Omega} \left| T_E(t)\right|^2.
\end{equation}

We calculate this quantity numerically and add that  to the quantity $CV^2$ dissipated in the switching circuit  to 
find the total dissipation $E_{total}$. The results 
are plotted as a function of $\tau$ in Fig. 2 of the main letter.

\section{Simulation results}
Some additional simulation results and corresponding discussions are given in the 
Figures~\ref{fig:energy_harvesting_theta_dynamics} -~\ref{fig:energy_harvesting_potential_profile}.

\clearpage
\pagebreak
\begin{figure}
\centering
\includegraphics[width=6in]{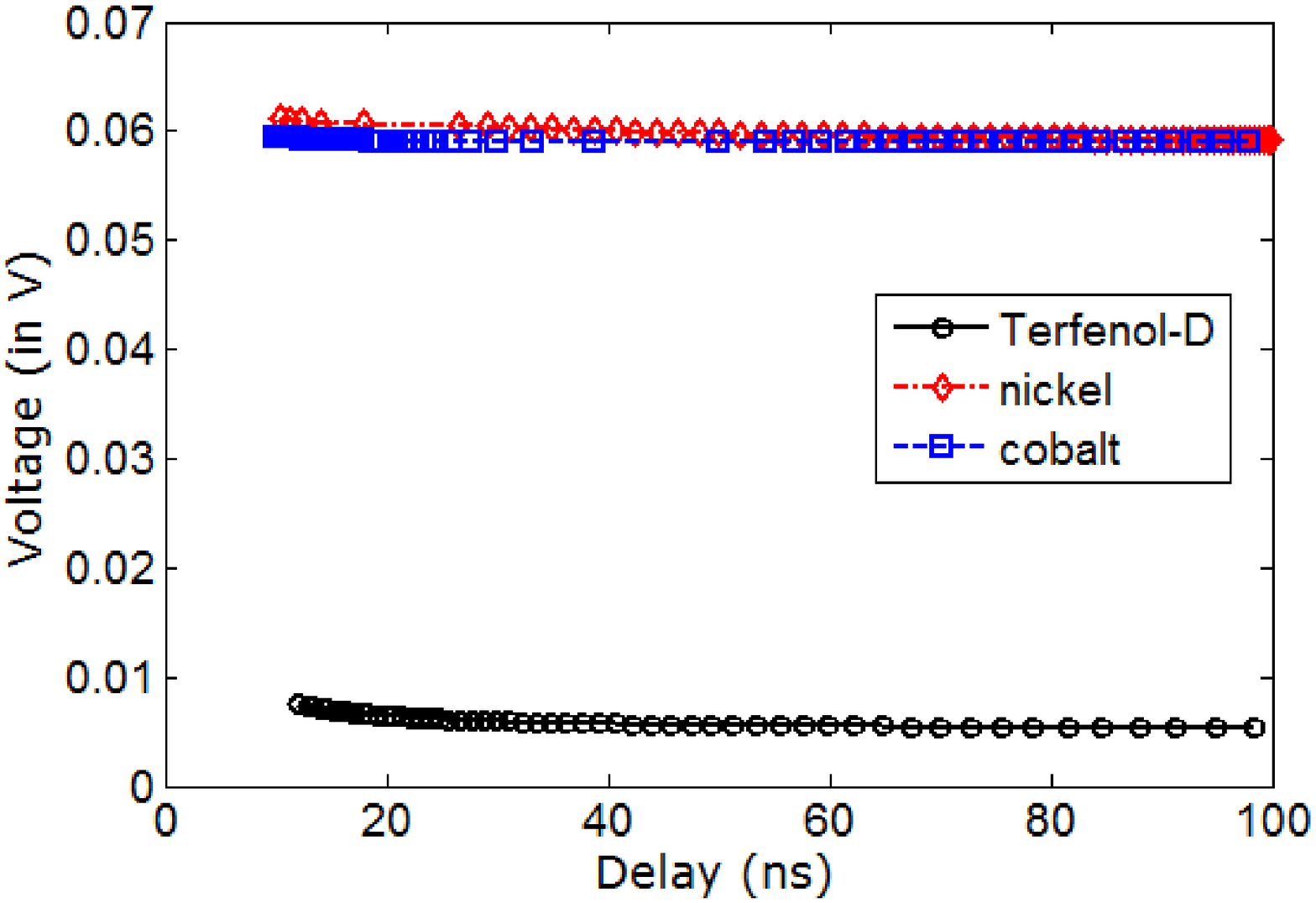}
\caption{\label{fig:energy_harvesting_delay_voltage} Voltage required to switch a multiferroic nanomagnet of the shape and size considered in this paper versus switching delay. Three different layers are considered for the magnetostrictive layer. Terfenol-D requires the smallest voltage since it has the highest magnetostrictive coefficient. This tiny voltage requirement makes this mode of switching magnets extremely energy-efficient.}
\end{figure}

\clearpage
\pagebreak
\begin{figure}
\centering
\includegraphics[width=6in]{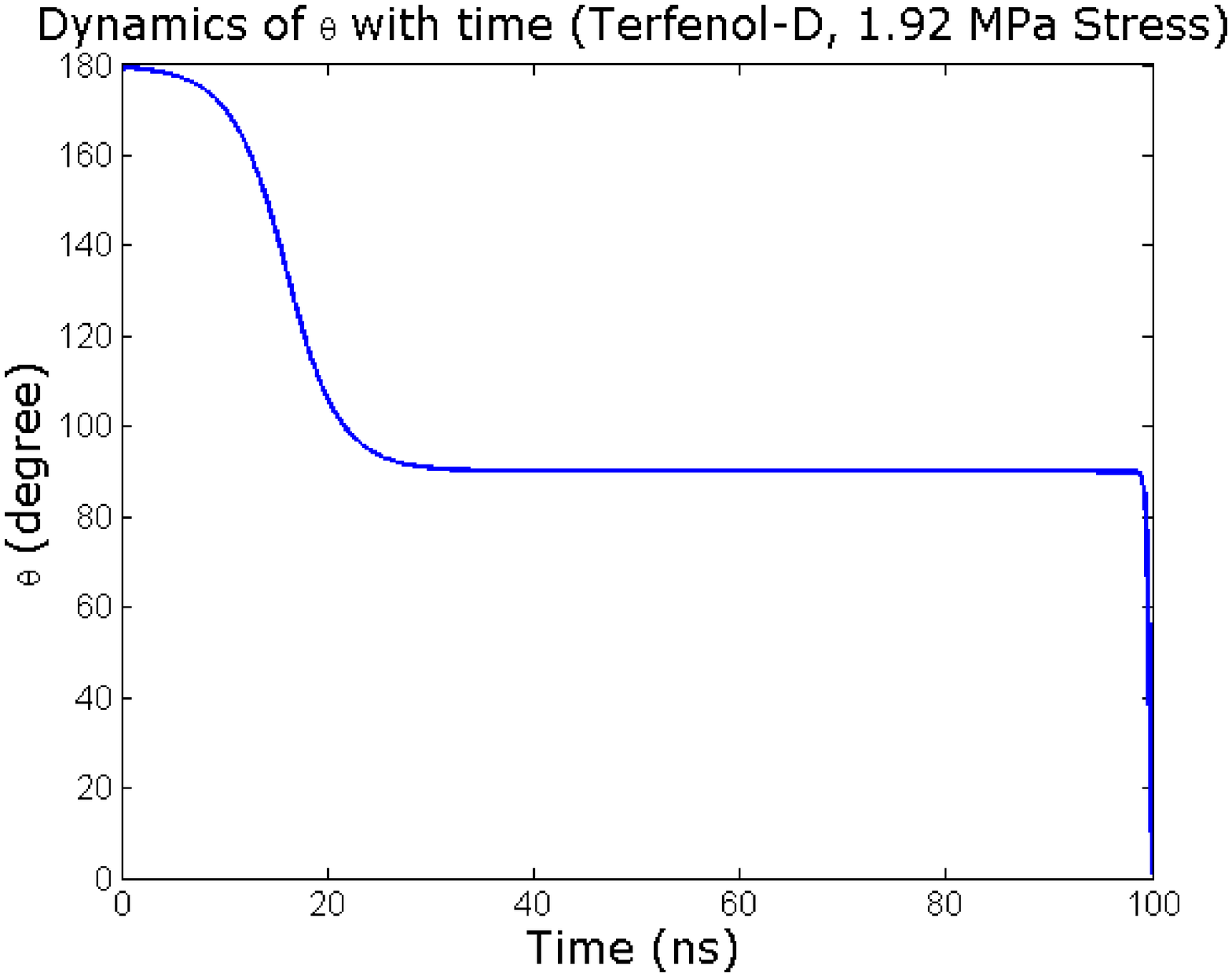}
\caption{\label{fig:energy_harvesting_theta_dynamics} Time evolution of the polar angle $\theta$ in a 
Terfenol-D/PZT multiferroic nanomagnet under 1.92 MPa applied stress. Stress is applied abruptly
at time $t$ = 0
to rotate the
magnetization away from its initial orientation close to the --z-axis ($\theta = 179^{\circ}$)
and it is removed abruptly once $\theta$ reaches 90$^{\circ}$, which corresponds approximately
to the hard axis. Thereafter, the magnetization spontaneously decays to the easy axis 
 since shape anisotropy
prefers the unstressed magnet's magnetization to align along the easy axis. Whether it decays 
to the +z-axis or --z-axis  is determined by the sign of $B_{0e}$ when $\theta$ reached 
90$^{\circ}$. If the sign is positive, then $\theta'$ is negative and $\theta$ will decrease with time, finally
reaching the value of 1$^{\circ}$ so that the magnetization aligns along the desired +z-axis.
It is therefore imperative to ensure that $B_{0e}$ is positive, which will happen only 
if $\phi < 90^{\circ}$ when $\theta$ = 90$^{\circ}$. We show in the next figure that this indeed happens as a consequence of the 
coupled $\theta$- and $\phi$-dynamics. The coupled dynamics therefore plays a critical role to ensure 
correct switching.
Note that the magnetization spends a lot of time 
around $\theta$ = 90$^{\circ}$ which is the hard axis. Once the magnetization gets past the hard axis, it
quickly reaches the easy axis. This can be understood by looking at the 
energy profile in Fig. \ref{fig:energy_harvesting_potential_profile}. The small stress causes a 
{\it shallow} energy minimum at the hard axis, but upon removal of stress, shape anisotropy 
causes a {\it tall} energy barrier at the hard axis. Hence it is much easier to approach the easy
axis from the hard axis, but much harder to approach the hard axis from the easy axis.
The total switching delay in this case is $\sim$ 100 ns, out of which nearly $\sim$90 ns is spent to get past the 
hard axis starting out from the easy axis, and $\sim$10 ns to decay to the easy axis from the 
hard axis.}
\end{figure}

\clearpage
\pagebreak
\begin{figure}
\centering
\includegraphics[width=6in]{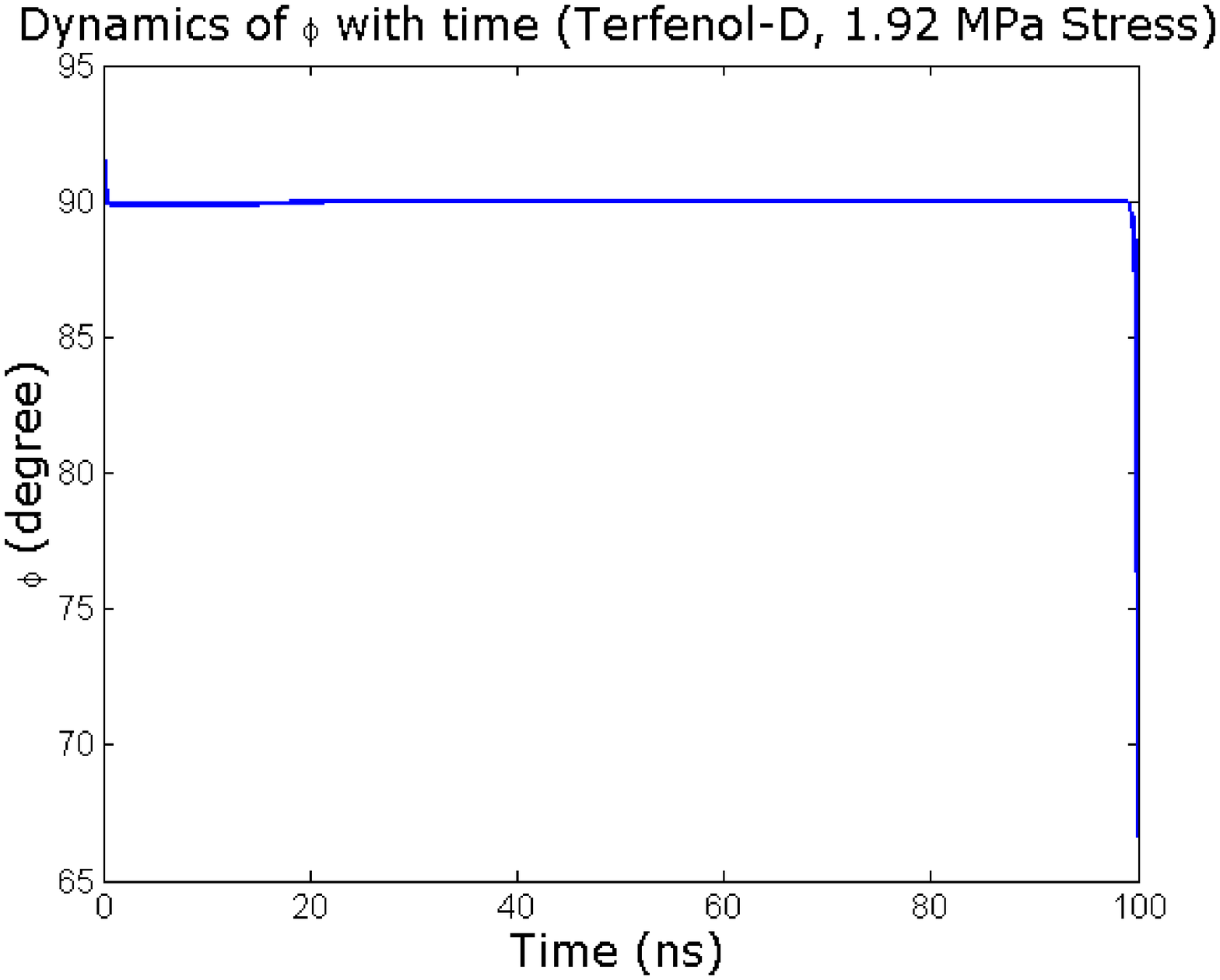}
\caption{\label{fig:energy_harvesting_phi_dynamics} Time evolution of the azimuthal angle $\phi$ 
during the switching of a 
Terfenol-D/PZT multiferroic nanomagnet subjected to 1.92 MPa stress. Initially, we start from 
$\phi=90^\circ$ and $\theta=179^\circ$ and therefore the second term in Equation~\eqref{eq:phi_dynamics}, 
$B (\phi(t)) cos\,\theta$ starts out as positive (since $B(\phi(t))$ goes negative upon application of stress and $cos\,\theta$ is 
negative in the range $90^\circ < \theta < 180^\circ$). Consequently $\phi$ decreases with time 
initially. This decrease of $\phi$ affects the $B_{0e}(\phi(t))$ term in 
Equation~\eqref{eq:theta_dynamics} facilitating the rotation of the magnetization angle $\theta$ 
from 179$^{\circ}$ toward $90^\circ$. Thereafter,  $\phi$ starts to increase because the term 
$B_{0e}(\phi(t))$ 
becomes non-zero as soon as $\phi$ deviates from $90^\circ$. But $\phi$ never reaches  exactly
 $90^\circ$ when $\theta$ = $90^\circ$. This avoids a possible stagnation point at the hard axis.
When $\theta=90^\circ$ [hard axis], stress is removed but the
finite value of $B_{0e}(\phi(t))$ [the term due to shape anisotropy]
continues to rotate the magnetization towards the +z-axis. 
When $\theta < 90^\circ$, the term 
$B (\phi(t))cos\,\theta$ goes positive and according to Equation~\eqref{eq:phi_dynamics}, 
$\phi$ starts to decrease. As $\phi$ decreases, $B_{0e}(\phi(t))$ increases and according to 
Equation~\eqref{eq:theta_dynamics}, $\theta$ decreases sharply and ultimately reaches a value of
1$^{\circ}$, at which point the switching is complete. Note that $\phi$ never deviates too far from
90$^{\circ}$ at this low value of stress (except at the very tail end of the switching), meaning that the 
magnetization vector is pretty much confined 
to the plane of the magnet (y-z plane) and its out-of-plane excursion is very small. Under high stresses, 
the out-of-plane excursion can be quite significant.}
\end{figure}

\clearpage
\pagebreak
\begin{figure}
\centering
\includegraphics[width=6in]{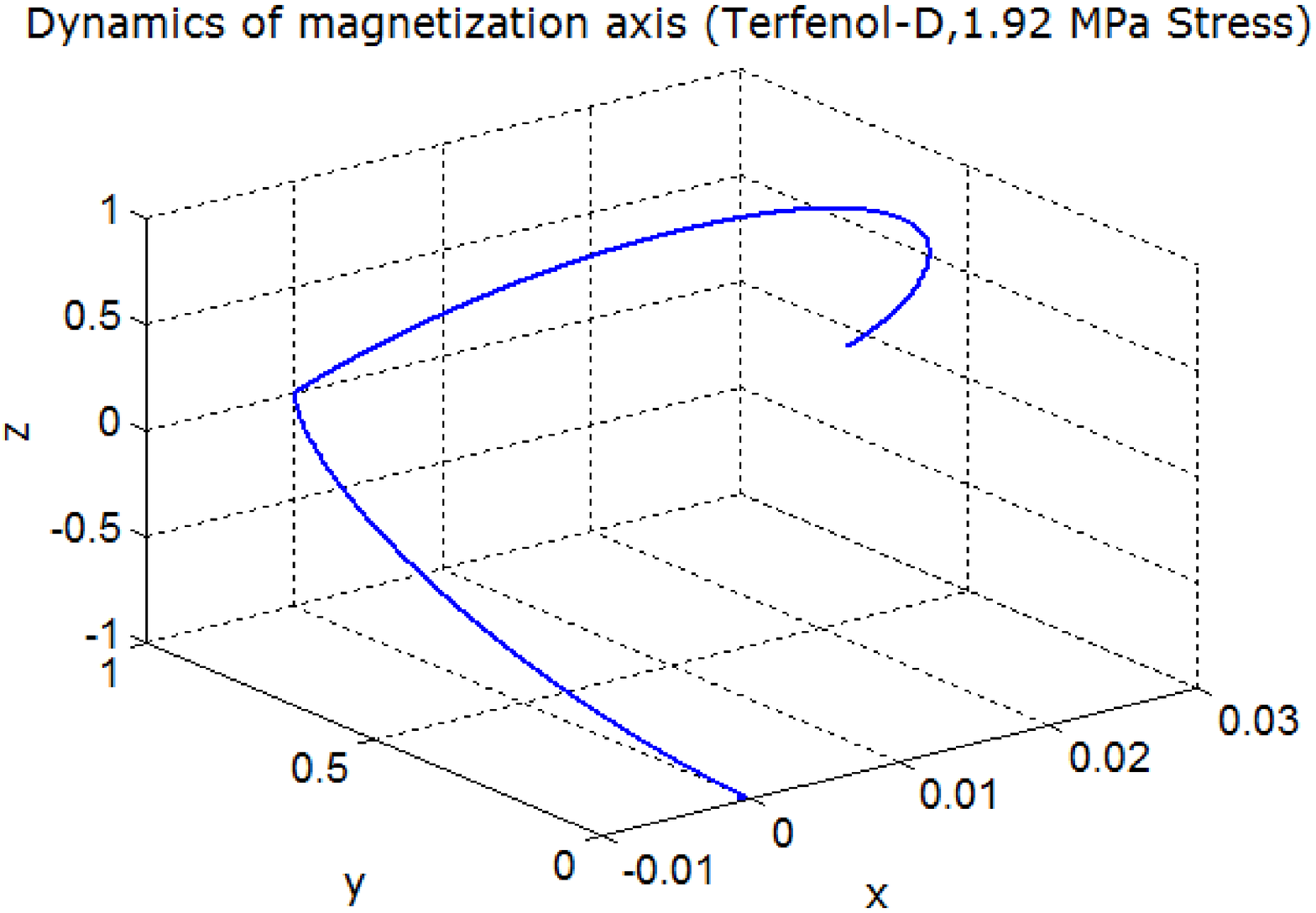}
\caption{\label{fig:energy_harvesting_magnetization_rotation} Trajectory traced out by the tip of 
the magnetization vector in space during switching. The magnet is a Terfenol-D/PZT multiferroic 
nanomagnet subjected to 1.92 MPa stress.}
\end{figure}

\clearpage
\pagebreak
\begin{figure}
\centering
\includegraphics[width=6in]{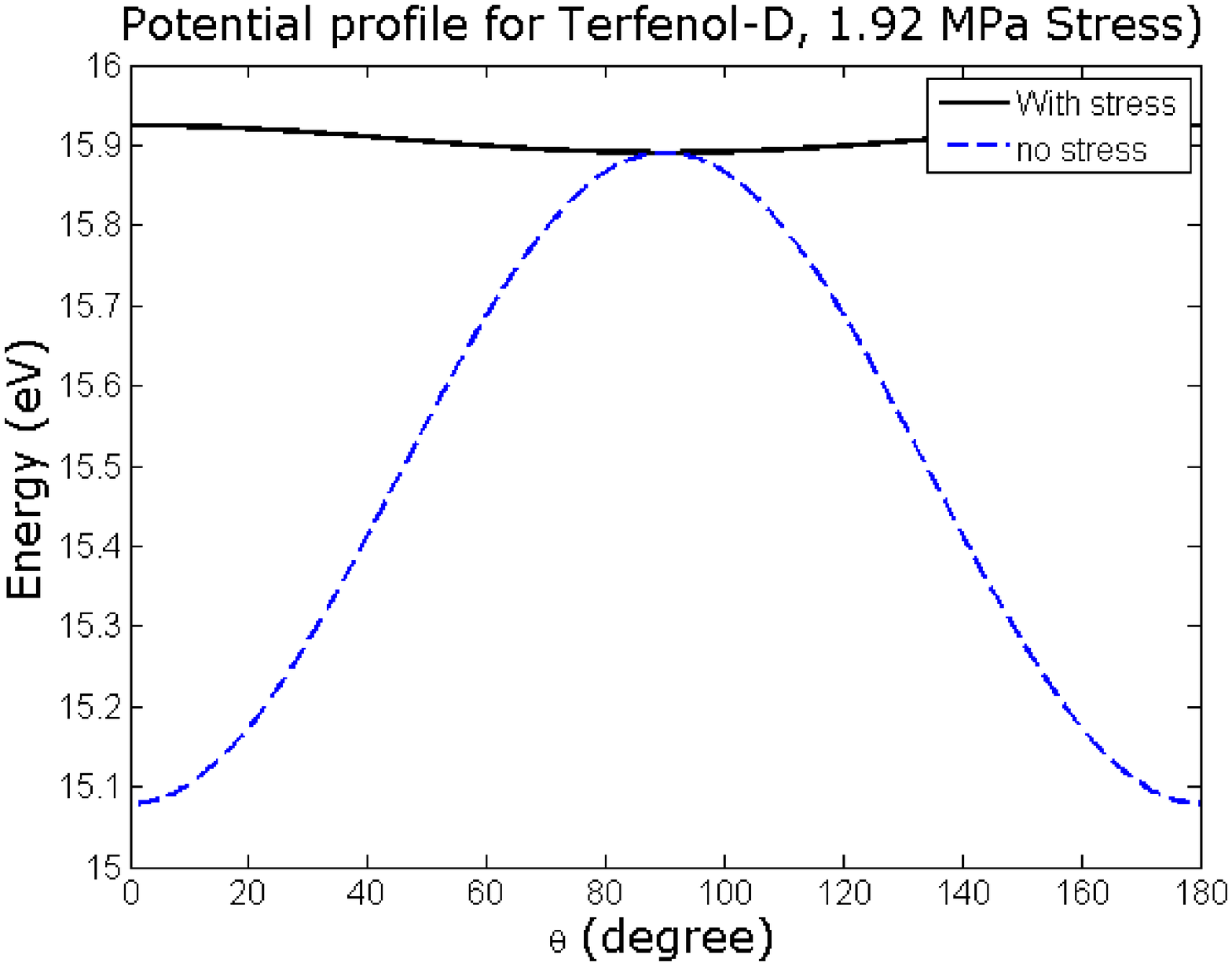}
\caption{\label{fig:energy_harvesting_potential_profile} Steady state energy profiles of a stressed 
and unstressed Terfenol-D/PZT multiferroic nanomagnet. The magnitude of the stress is
1.92 MPa. Without any applied stress, the potential profile depicts the shape anisotropy energy barrier
which is 0.8 eV or 32 kT.}
\end{figure}

\clearpage
\pagebreak

\makeatletter 
\renewcommand\@biblabel[1]{[S#1]}
\makeatother
